\documentclass{aa}
\usepackage{graphicx}
\usepackage{txfonts}
\usepackage{amsmath}	
\usepackage{amssymb}	
\usepackage{caption}
\usepackage{subcaption}

\newcommand{\Msun}{M_\odot}
\newcommand{\Mdot}{\dot{M}}
\newcommand{\Mdotstar}{\dot{M}_\ast}
\newcommand{\Mdotin}{\dot{M}_\mathrm{in}}

\newcommand{\Mdotacc}{\dot{M}_\mathrm{acc}}

\newcommand{\Pdot}{\dot{P}}

\newcommand{\rin}{r_\mathrm{in}}

\newcommand{\rco}{r_\mathrm{co}}
\newcommand{\rinmax}{r_\mathrm{in,max}}
\newcommand{\rA}{r_\mathrm{A}}
\newcommand{\reta}{r_\eta}
\newcommand{\rxi}{r_\xi}

\newcommand{\Rinmax}{R_\mathrm{in,max}}
\newcommand{\DeltaR}{\Delta r/r_\mathrm{in}}

\newcommand{\vesc}{v_\mathrm{esc}}

\newcommand{\Ostar}{\Omega_\ast}
\newcommand{\OmegaK}{\Omega_\mathrm{K}}

\newcommand{\Lx}{L_\mathrm{X}}

\newcommand{\Gammaacc}{\Gamma_\mathrm{acc}}
\newcommand{\GammaD}{\Gamma_\mathrm{D}}
\newcommand{\Gammadip}{\Gamma_\mathrm{dip}}

\newcommand{\tint}{t_\mathrm{int}}

\newcommand{\gpers}{g~s$^{-1}$}
\newcommand{\ergpers}{erg~s$^{-1}$}

\newcommand{\spers}{s~s$^{-1}$}

\newcommand{\Alfven}{Alfv$\acute{\mathrm{e}}$n~}
\newcommand{\src}{\object{4U 1626–67}}
\newcommand{\Fx}{F_\mathrm{x,bol}}

\DeclareUnicodeCharacter{2212}{-}

\begin{document}

   \title{On the torque reversals of \src}


   \author{A. A. Gen\c{c}ali
           \inst{1}
           \and
           N. Niang
           \inst{1}
           \and
           O. Toyran
           \inst{1}
           \and
           \"{U}. Ertan
           \inst{1}
           \and
           A. Ulubay
           \inst{2}
           \and
           S. {\c S}a{\c s}maz
           \inst{3}
           \and
           E. Devlen
           \inst{4}
           \and\\
           A. Vahdat
           \inst{5}
           \and
           {\c S}. {\" O}zcan
           \inst{6}
           \and
           M. A. Alpar
           \inst{1}
          }

   \institute{Sabanc{\i} University, Orhanl{\i}, Tuzla, 34956, \.{I}stanbul, Turkey\\
              \email{gencali@sabanciuniv.edu}
         \and
             Faculty of Science, Department of Physics, \.{I}stanbul University, 34134, Vezneciler, \.{I}stanbul, Turkey
         \and    
             Istanbul Technical University, Faculty  of Science  and  Letters, Physics Engineering Department, 34469, \.{I}stanbul, Turkey 
         \and
             Faculty of Science, Department of Astronomy and Space Sciences, Ege University, 35100, Bornova, \.{I}zmir, Turkey
         \and
             Institut f\"ur Astronomie und Astrophysik, Universit\"at Tübingen, Sand 1, D-72076 T\"ubingen, Germany
         \and         
             Cumhuriyet mah., Z{\" u}mr{\" u}t sok., {\" U}sk{\" u}dar, 34699, \.{I}stanbul, Turkey
             }

   \date{Received July 12, 2021; accepted October 25, 2021}

 
  \abstract
   {We have investigated the detailed torque-reversal behavior of \src~in the framework of the recently developed comprehensive model of the inner disk radius and torque calculations for neutron stars accreting from geometrically thin disks. The model can reproduce the torque -- X-ray luminosity relation across the torque reversals of \src. Our results imply that: (1) rotational equilibrium is reached when the inner disk radius equals the co-rotation radius, $\rco$, while the conventional \Alfven radius is greater than and close to $\rco$, (2) both spin-up and spin-down torques are operating on either side of torque reversal, (3) with increasing accretion rate the spin-up torque associated with accretion onto the star gradually dominates the spin-down torque exerted by the disk. The torque reversals are the natural outcome of transitions between the well-defined weak-propeller and spin-up phases of the star with a stable geometrically thin accretion disk.}

   \keywords{accretion -- accretion disks – pulsars: individual: (4U 1626--67) }

   \maketitle
%

\section{Introduction}

Neutron stars in low-mass X-ray binaries (LMXBs) receive mass from their companion through Roche-lobe overflow. The interaction between the magnetosphere and the accretion disk, and the mass accretion onto the neutron star govern the X-ray luminosity and the rotational behavior of the star \citep[for a review of disk accretion onto magnetic stars see][]{Lai2014}. Comparing observations of these systems with theoretical models, it is estimated that there are mainly three rotational phases of neutron stars in LMXBs, namely, the strong-propeller (SP), the weak-propeller (WP), and the spin-up (SU) phases. In the SP phase, all the inflowing mass is expelled from the inner boundary of the disk by the magnetic torques, and the star spins down. In the WP phase, most of the matter from the inner disk is accreted onto the star, while the star still slows down. In the SU phase, all the matter arriving at the inner disk flows onto the star along the closed magnetic field lines, and the star spins up \citep{Bildsten1997,Papitto2020}.

Accretion onto the star takes place when the inner disk radius, $\rin$, is equal to or smaller than the co-rotation radius, $ \rco$, at which the angular frequency of the neutron star, $\Ostar$, equals the Keplerian angular speed, $\OmegaK$, of the disk matter. In the models, for a given disk mass-flow rate, $\Mdotin$, and magnetic dipole moment, $\mu$, almost all the rotational and accompanying radiative properties of the star depend on the location of $\rin$. Equating the magnetic and viscous stresses, $\rin$ was conventionally estimated to be close to the conventional \Alfven radius, $\rA~\simeq~\Big[ \mu^4/(GM~\Mdotin^2) \Big] ^{1/7}$, where $G$ is the gravitational constant, and $M$ is the mass of the star \citep{Davidson1973, Lamb1973}. With $\rin = \rxi = \xi \rA$ and  $0.5 \leq \xi \leq 1$ \citep{Illarionov1975, Ghosh1979b, Arons1993, Ostriker1995}, these models encounter the following difficulties in explaining the characteristic X-ray luminosity, $\Lx$, and rotational behavior of neutron stars in LMXBs \citep{Bildsten1997, Archibald2009, Papitto2013, Bassa2014, Jaodand2016, Papitto2020}: (1) There is ongoing accretion with rates much smaller than the rate corresponding to the $\rA = \rco$ condition. (2) For a large range of accretion rates $\Mdotin$ (up to 2 orders of magnitude depending on the $\xi$, $\eta$, $\mu$ and $P$ parameters described in Sect. \ref{model}), spin-down continues while there is mass accretion onto the star. (3) The transition to the SP phase takes place at very low $\Mdotin$ when $\rA \gg \rco$ \citep{Bassa2014, Archibald2015, Papitto2015, Papitto2020}. (4) Transitions occur between the spin-up and spin-down regimes with similar torque magnitudes without a significant change in $\Lx$ \citep{Deeter1989, Bildsten1997, Chakrabarty1997a, Chakrabarty1997b, Inam2009, Camero-Arranz2010, Takagi2016}.

In the early detailed models of disk-field interaction, accretion disks were assumed to  be threaded by closed magnetic field lines of the star \citep{Ghosh1979a,Ghosh1979b}. Later, \citet{Wang1987} showed that this picture is not self-consistent, because winding of the toroidal field lines resulting from the disk-field interaction would destroy most of the disk. Magnetically threaded disk models also have difficulties in explaining the abrupt torque reversals corresponding to small changes in the accretion rate. A series of work showed that the field lines interacting with the inner disk tend to inflate and open up when the toroidal field strength becomes comparable to that of the poloidal field \citep{Aly1990, vanBallegooijen1994,Lynden-Bell1994,Lovelace1995, Uzdensky2002}. In the model proposed by \citet{Lovelace1995} this result was found to be applicable to the propeller phase as well \citep{Lovelace1999, Ustyugova2006}. In this model the field lines and the inner disk interact in a narrow boundary region with ongoing opening and reconnection cycles of the magnetic field lines on dynamical timescales, while the field lines are decoupled from the disk outside the boundary region (see Sect. \ref{model}). This model was supported by the results of the numerical simulations \citep{Hayashi1996, Goodson1997, Miller1997}. 

Adopting the basic principles of the model proposed by \citet[][see Sect. \ref{model}]{Lovelace1995}, it was shown through analytic calculations within a new torque model that a steady SP phase can be established with $\rin$ much smaller than $\rA$ \citep{Ertan2017, Ertan2018}. Applying this model to the transitional millisecond pulsars (tMSP), the critical $\Mdotin$ level for the SP/WP transition was found to be consistent with the estimated accretion rates and the torque variations corresponding to transitions between the radio pulsar and X-ray pulsar states \citep{Archibald2009, Papitto2013, Bassa2014, Jaodand2016, Papitto2020}. The extended version of this model, including the SU phase as well, can also account for the general torque-reversal characteristics summarized above \citep{Ertan2021}. The model estimates of $\rin$ are seen to be much smaller than the conventional $\rA$ in the SU phase as well. Only for a narrow range of $\Mdotin$, around the torque-reversal, $\rin$ traces $\rxi$ \citep[for details see][]{Ertan2021}. 

In this work, we investigate the detailed torque-reversal behavior of \src~in the model proposed by \cite{Ertan2021}. \src~seems to be the best source for testing the thin disk torque-reversal models \citep[for a detailed discussion see][]{Camero-Arranz2010}. Because, the source shows torque reversals purely by disk torques due to the lack of wind accretion from the companion. In addition, there are many period derivative and X-ray luminosity measurements available on either side of the torque reversal \citep{Camero-Arranz2010, Takagi2016}. \src~was first discovered with Uhuru in 1972 \citep{Giaconni1972} in a $42$~min ultracompact binary system, accreting matter from a companion with mass $ M\sim 0.04~\Msun$, with some uncertainties depending on the inclination angle of the LMXB \citep{Middleditch1981, Levine1988, Chakrabarty1997a, Chakrabarty1998}. With a spin period of $P = 7.66$~s \citep{Rappaport1977}, the source shows torque reversals without a significant change in its X-ray luminosity \citep[$\Lx \sim 10^{36}~-~10^{37}$~\ergpers;][]{Chakrabarty1997a, Camero-Arranz2010}. The distance is estimated to be in the $5~-~13$~kpc range \citep{Chakrabarty1998, Takagi2016}. Magnitudes of the measured period derivatives are in the range $1.8\times 10^{-12} < |\Pdot| < 1.3\times 10^{-10}$~\spers~ \citep[][and references therein]{Takagi2016}. 

Detection of double-peaked emission lines in the X-ray band indicates the presence of an accretion disk \citep{Schulz2001}. A cyclotron line detected at $\sim 37$~keV corresponds to a surface magnetic field strength $B \sim 3 \times 10^{12}$~G \citep{Orlandini1998, Coburn2002, Dai2017}. While the X-ray flux increased by a factor of $\sim 3$ from 2006 to 2010, no concurrent changes in the magnetic field strength, as inferred from cyclotron line energies, accompanied the flux changes and associated torque reversals \citep{Camero-Arranz2012}. The range of inner disk radii recently estimated for \src~from the double-peaked emission lines observed with Chandra/LETGS \citep{Hemphill2021} indicates that $\rin \simeq \rco$ for plausible neutron star masses within the uncertainty in the disk inclination angle. During the torque reversals, the shape of the pulse profile changes, which could be a sign of a change in the emission geometry of the accretion column \citep{Beri2014, Koliopanos2016}. 
 
 We summarize the model \citep{Ertan2021} in Sect. \ref{model}, and discuss our results together with a summary of other torque-reversal models in Sect. \ref{app}. Our conclusions are summarized in Sect. \ref{conc}.

\section{The model}
\label{model}


For the inner disk radius, $\rin$, and the torque calculations, we use the model developed earlier for neutron stars accreting from geometrically thin accretion disks \citep[for details, see][]{Ertan2017,Ertan2018,Ertan2021}. Here, we describe the model briefly. The inner disk interacts with the magnetic field lines in a boundary layer between $\rin$ and $\rin + \Delta r$. In this region, the closed field lines cannot slip through the disk since the interaction time-scale, $\tint \simeq |\Ostar − \OmegaK|^{−1}$, is orders of magnitude shorter than the diffusion time-scale of the magnetic field lines \citep{Fromang2009}. In this boundary region, field lines inflate and open up within $\tint$. In the SP phase, matter can be expelled from the inner disk along the open field lines. The open field lines reconnect on a dynamical timescale comparable to $\tint$, completing the cycle. The field lines are decoupled from the disk at radii greater than $\rin + \Delta r$ \citep{Lovelace1999, Ustyugova2006}.

A strong-propeller mechanism can be sustained at $\rin > \rco = (G M / \Ostar^{2})^{1/3}$, provided that 
 the field lines can throw out all the matter flowing into the boundary region. This requires that the field lines are able to force the matter into co-rotation in the interaction boundary. The maximum inner disk radius, $\rinmax$, at which this strong-propeller mechanism is satisfied, is determined by the implicit condition
\begin{equation}	
	\Rinmax^{25/8}~|1 - \Rinmax^{-3/2}| ~ \simeq ~ 1.26  ~
\alpha_{-1}^{2/5} ~M_{1.4}^{-7/6} ~\Mdot_\mathrm{in,16}^{-7/20}~ \mu_{30} ~ P^{-13/12}
\label{eq:Rin} 	
\end{equation}
\vspace{3mm}
 \citep{Ertan2017} where $\Rinmax = \rinmax/ \rco$, and $\alpha_{-1} = (\alpha /0.1)$ is the kinematic viscosity parameter, $M_{1.4} = (M / 1.4 \Msun )$, $\Mdot_\mathrm{in,16}~=~\Mdotin/(10^{16}$~\gpers), $\mu_{30} = \mu / (10^{30}$~G~cm$^3$), and $P$ is the rotational period of the star. We define $\reta \equiv \eta~\rinmax$ where $\eta~\leq~1$ is likely to be close to unity because of the sharp variation of magnetic torques with radial distance. In the SP phase, $\rin = \reta$ is always found to be much smaller than $\rxi = \xi \rA$. For a given dipole field strength, the dependence of $\rin$ on $\Mdotin$ is much weaker than that of $\rA \propto \Mdot^{-2/7}$ in the SP phase.

The speed of the closed field lines rotating with the star is greater than the escape speed, $\vesc$, outside the radius $r_1~=~1.26~\rco$, while it is smaller than $\vesc$ between $\rco$ and $r_1$. A steady-state SP phase is possible only with $\rin > r_1$. In the case that $\rin$ is instantaneously between $\rco$ and $r_1$, the field lines can effectively throw the matter from the inner disk, while the matter returns back to the disk at larger radii. Due to growing pile-up of this matter, the inner disk moves inwards down to $\rin = \rco$ eventually switching accretion on, and the system enters the WP phase. The SP/WP transition takes place when $\rin~=~\reta~=~r_1$, and $\rxi$ is much greater than $\reta$ and $\rco$. In the WP phase, the inner disk cannot penetrate inside $\rco$, while matter flows to the star from $\rco$. For the inner disk to protrude inside $\rco$ the viscous stresses should dominate the magnetic stresses at this radius. With increasing $\Mdotin$, $\rxi$ approaches $\rco$ in the WP phase. The inner disk starts to propagate inwards of $\rco$ when $\Mdotin$ exceeds the critical rate corresponding to $\rxi = \rco$ \citep[for details see][]{Ertan2021}. The torque reversal takes place at a certain $\Mdotin$ when $\rin = \rco$, and $\rxi$ is close to $\rco$ depending on the actual value of $\xi$. 

In the SU phase, with the solution for $\reta < \rco$ beyond a certain $\Mdotin$ level, a steady $\rin$ cannot be established at $\rxi$. The inner disk extends inwards opening the field lines until $\rin = \reta$, the stable inner disk radius corresponding to the same accretion rate $\Mdotstar$. For higher accretion rates, $\rin$ tracks $\reta$ which could be several times smaller than $\rxi$ depending on the magnetic dipole moment and the period of the source. Considering that the field lines are decoupled from the disk for $r > \rin + \Delta r$, we estimate that the spin-down torque, $\GammaD$, switches off when $\rin$ enters inside $\rco$ by a small radial extent comparable to $\Delta r$ \citep[for a detailed description see][]{Ertan2021}.

The total torque acting on the star can be written as
\begin{equation}
	\Gamma = - ~\frac {\mu^2}{\rin^3} \left(\frac{\Delta r}{\rin}\right) ~+~ \sqrt{G M \rin} ~\Mdotstar + \Gammadip
\label{eq:torque}
\end{equation}	
where the first term is the spin-down torque, $\GammaD$, produced by the interaction between the inner disk and the magnetic field of the star in the narrow boundary region with radial width $\Delta r < r$. The second term is the spin-up torque, $\Gammaacc$, associated with the mass accretion from the inner disk onto the star \citep{Pringle1972}. Note that $\Mdotstar$ is the accretion rate onto the star, while $\Mdotin$ denotes mass inflow-rate of the disk. The magnetic dipole torque, $\Gammadip \simeq -2 \mu^2 \Ostar^3 / 3c^3$ where $c$ is the speed of light, also contributes to the spin-down torque, and is in most cases negligible in comparison with $\GammaD$. In the SP phase ($\rin = \reta > r_1$), there is no mass accretion onto the star and thus $\Gammaacc = 0$. In the WP phase, both spin-up and spin-down torques are active with $\rin = \rco$. In this phase, with increasing accretion rate, $|\Gammaacc|$ increases, while $|\GammaD|$ remains constant, since $\rin = \rco$ persists for a large $\Mdotstar$ range in the WP phase. Eventually, at a certain $\Mdot_\ast$, the spin-up and spin-down torques are balanced, and the system makes a transition to the SU phase (torque reversal). We note that the torque reversal does not necessarily take place when $\rxi$ is exactly equal to $\rco$. Nevertheless, our results indicate that these two radii are close to each other during the torque reversals. With increasing $\Mdotin$, $\rin$ enters inside $\rco$ tracking $\rxi$ for a narrow range of $\Mdotstar$. Outside this $\Mdotin$ range, $\rin$ is significantly smaller than $\rxi$, depending on $B$, $P$ and $\Mdotin$ \citep[for a detailed description see][]{Ertan2021}.

    \begin{figure}
    \centering
        \begin{subfigure}[b]{0.5\textwidth}
         \centering
        \includegraphics[width=\textwidth]{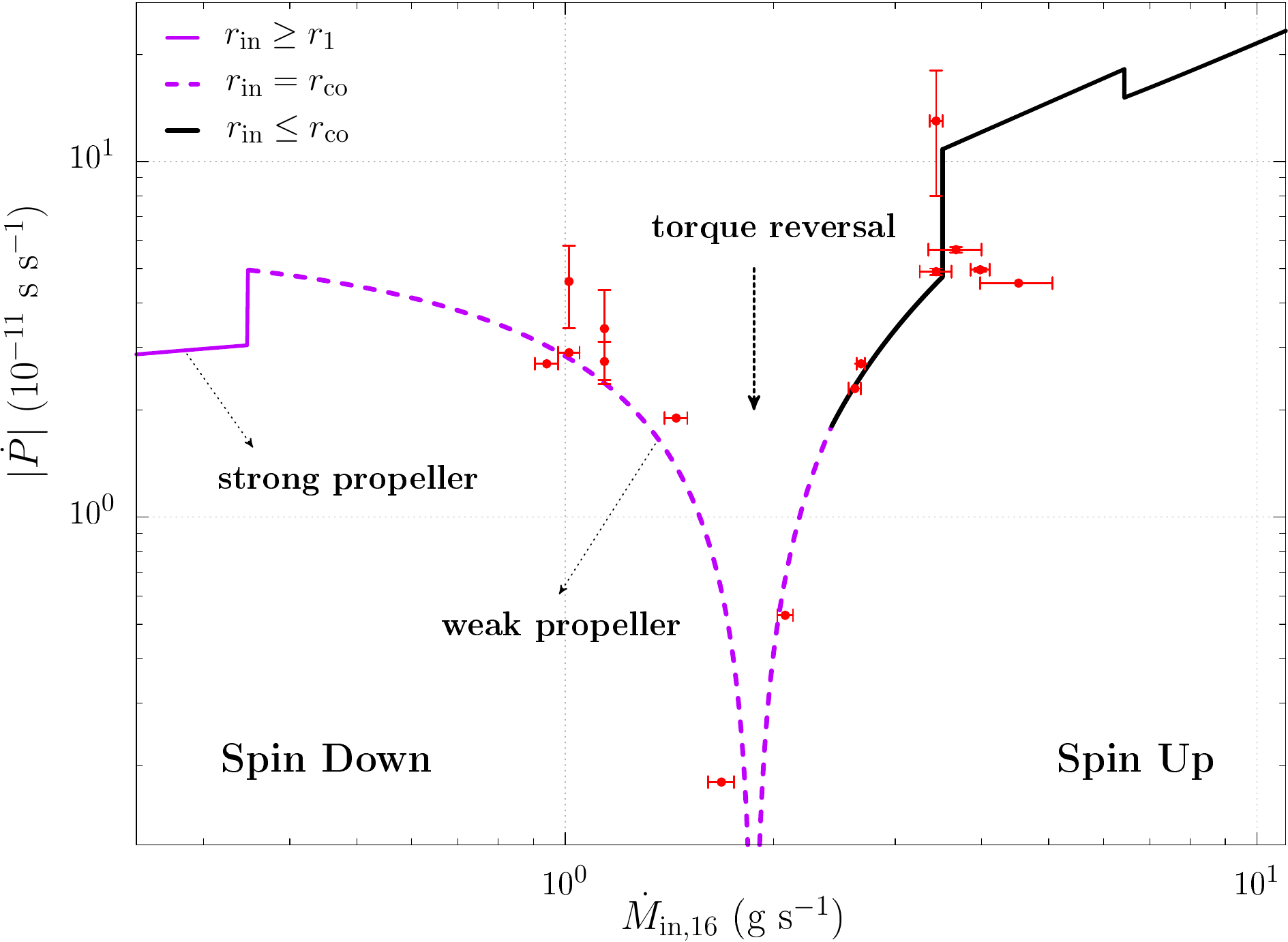}
        \caption{$\xi = 0.68$ \vspace{0.2cm}}
        \label{fig:2a}
    \end{subfigure}
     \hfill
    \begin{subfigure}[b]{0.5\textwidth}
    \centering
        \includegraphics[width=\textwidth]{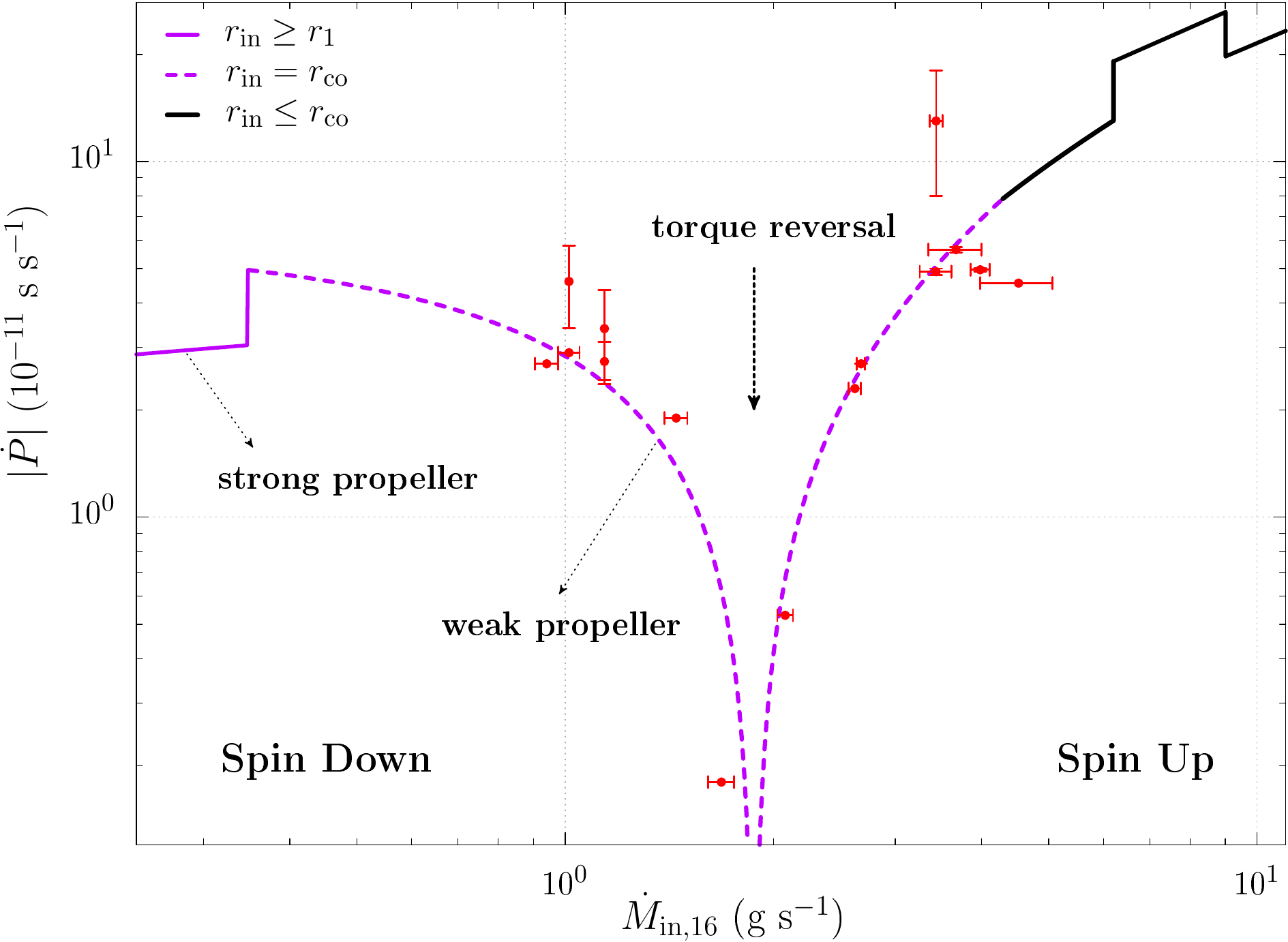}
        \caption{ $\xi~=~0.80$}
        \label{fig:2b}
        \end{subfigure}
        \caption{The torque reversal of \src~for different $\xi$ values. The model parameters are $B~=~3~\times~10^{12}$~G, $\DeltaR~=0.2~$, $\eta~=1~$, $d~=~5.3$~kpc. We obtained the data from \citet[][and references therein]{Takagi2016}. For both models, the spin-down torque is inactivated when $\rin~=~\rco~-~\Delta~r/2$ (see the text for details). }
        \label{fig2}
    \end{figure}

\section{Application to \src~and discussion} 
\label{app}

\begin{figure}
	    \centering
    	\includegraphics[width=1\columnwidth]{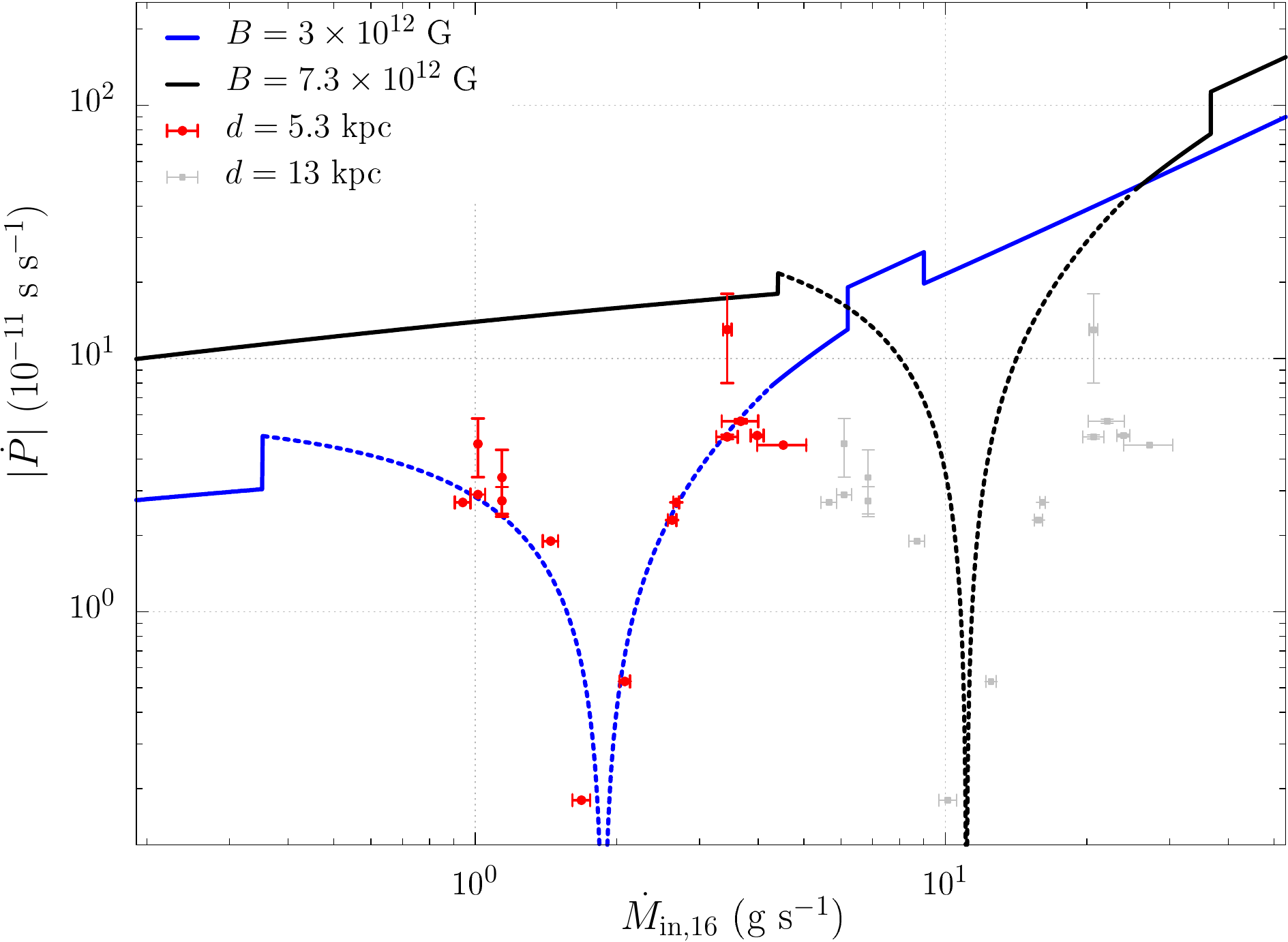}
        \caption{Illustrative model curves with torque-reversals corresponding to $\Mdotin$ values. The blue model curve is the same as given in Fig. \ref{fig:2b} ($\DeltaR~=0.2~$, $\eta~=1~$ and $\xi~=~0.8$). The black curve is also obtained with the same disk parameters, but for a larger distance that requires a stronger $B$ and a higher $\Mdotin$ during the torque reversal. The magnetic field $B$ and distance $d$ values used in the models are given in the figure. It is clearly seen that the model with the larger $d$ cannot reproduce the data (grey data points).}
        \label{fig3}
    \end{figure}

We take the measured $\Pdot$ and bolometric X-ray flux, $\Fx$, values of the source from \citet[][and references therein]{Takagi2016}. The distance, $d$, estimates range from $5$ to $13$~kpc \citep{Chakrabarty1998, Takagi2016}. Our model calculations are sensitive to $\Mdotstar$, and thus to distance. $\Fx$ values are converted to bolometric X-ray luminosity and $\Mdotstar$ using $\Lx~=~4~\pi~d^{2}~\Fx~=~GM\Mdotstar/R_\ast$ neglecting beaming effects. The beaming correction was estimated to be at most $50\%$ for Her X-1 \citep{Basko1975}. Due to similarity of the two sources, \citet{Takagi2016} estimated a similar correction for \src. In this case, we obtain reasonable model fits with somewhat larger distances. For instance, if the actual beaming correction is $50\%$, the same model fit given in Fig. \ref{fig2} is obtained with a $\sim 40\%$ larger distance ($\sim 7.5$~kpc), remaining inside the estimated distance range. In our model, the relation between the torques and the accretion rates is obtained for steady-state disks. Torque reversals are short-term events, and during these transitions, there could be fluctuations in the local mass inflow rate, X-ray luminosity and structure of the inner disk that could deviate from the steady-state disk properties. Since these time-dependent fluctuations are not addressed in our model, we have looked for the seemingly best fits to the observed data without performing a $\chi^2$ test. We have obtained the model curves in Fig. \ref{fig2} tracing the critical $\Mdotstar$ values for the torque reversal and the corresponding $B$ values for the period of \src. It is seen in Fig. \ref{fig2} that the model produces the observed torque reversal of \src~quite well. Reasonable model curves are obtained with $B \simeq 3 \times 10^{12}$~G, and the torque-reversal takes place at $\Mdotstar \simeq 1.9\times 10^{16}$~\gpers~which corresponds to $d \simeq 5.3$~kpc. For larger distances,~$d$, which require greater $B$ and higher $\Mdotstar$ values during the torque reversal, the model cannot reproduce the observed $\Mdotin~-~\Pdot$ relation (see Fig. \ref{fig3}).

The critical $\Mdotin$ for the torque reversal is sensitive to the $\xi$ parameter. For $0.63 < \xi < 1$, the torque reversal takes place when $\rin = \rco$ and $\rxi \sog \rco$ with reasonable model fits. For instance, for the models seen in Fig.~\ref{fig2} the WP/SU transitions take place when, $\rxi\simeq~1.07~\rco$ for $\xi = 0.68$, and $\rxi\simeq~1.27~\rco$ for $\xi = 0.80$. For this range of $\xi$, the rotational equilibrium is obtained at the same $\Mdotin$. For $0.50 < \xi < 0.63$, $\rxi$ penetrates inside $\rco$ while the spin-down torque is still dominant. In this case, the spin-down torque is switched off, and the torque changes sign at a relatively low $\Mdotin$. This type of torque reversal gives reasonable model curves only for a very narrow range of $\xi$ ($0.60~-~0.63$) favoring $\rin = \rxi$ and $\rxi \sol \rco$ during the torque reversal. 

In Fig. \ref{fig2}, the sharp increase in $\Pdot$ in the spin-up regime is due to the inactivation of the spin-down torque, $\GammaD$, while the subsequent sharp decrease is produced by the rapid inward propagation of $\rin$ from $\rxi$ to $\reta$ as described in Sect. \ref{model}. Fig. \ref{fig2} illustrates how these features depend on the $\xi$ parameter. The abrupt rise in $\Pdot$ during the SP/WP transition corresponds to the inward motion of $\rin$ from $r_1$ to $\rco$, which results from the growing pile-up at the inner disk when $\rin$ gets into the range between $r_1 = 1.26~\rco$ and $\rco$ (See Sect. \ref{model}). The $\eta$ parameter affects the critical $\Mdotin$ for the SP/WP transition, while it does not change the critical $\Mdotin$ or the morphology of the torque reversal. With smaller $\eta$ values, the SP/WP transition takes place at lower $\Mdotin$ levels.

Illustrative model curves showing the dependence of torque-reversal characteristics on the model parameters were given by \citet{Ertan2021}. These model curves indicate that the morphologies of the torque reversals are likely to be similar for sources with very different dipole moments, periods and the accretion rates during the torque reversals, assuming that all these systems have similar $\xi$ and $\eta$ values, and are not affected by the winds of their companions. This is a testable prediction of the model in addition to the abrupt torque variations without significant changes in $\Lx$ in the SU phase (see Fig. \ref{fig2}). 

Recently, \citet{Benli2020} showed that the torque-reversal behavior of \src~can be accounted for with the assumption that $\rin$ remains equal to $\rco$ before and after the torque reversal without addressing the physical reason for this behavior and the condition for $\rin$ to penetrate inside $\rco$. Investigating the functional behavior of the torque reversal of the same source, \citet{Turkoglu2017} found that the torque variation as a function of $\Mdotstar$ contradicts the estimates of the conventional models assuming that the torque reversal occurs when $\rxi = \xi \rA = \rco$. 

There have been different proposals and models to explain the physical mechanism of the torque reversal. Transition of a warped disk into a retrograde regime was proposed as the reason for the spin-up/spin-down transition \citep{Makishima1988, Chakrabarty1997a, Nelson1997, Kerkwijk1998, Wijers1999, Beri2015}. The difficulty of this model is that there is evidence for the presence of a prograde disk in the spin-down phase  from the quasi-periodic oscillation analysis \citep{Kerkwijk1998}. Another idea is the variation of $\rin$ due to  transitions of the inner disk between the optically thin and thick regimes \citep{Vaughan1997, Yi1997,Dai2006, Yi1999}. In this model, the inner disk is Keplerian in the optically thick state, while the mass inflow is spherical-like and sub-Keplerian in the optically thin state. For \src, with a high accretion rate ( $\Mdotin > 10^{16}$~\gpers) and large radius ($\rin \sim 7 \times 10^8$~cm) during the transition, the inner disk is likely to be optically thick. For comparison, tMSPs are estimated to have optically thick disks in their sub-luminous ($\Lx \sim 10^{33}$~\ergpers) LMXB states with $\rin \sim 20$~km \citep{Papitto2020}. \citet{Lovelace1999} proposed that the spin-down torque is produced by a strong propeller mechanism throwing most of the inflowing matter out of the system when $\rin > \rco$. The propeller mechanism is switched off, accretion onto the star is allowed, and the star spins up when $\rin < \rco$. The transitions between the two phases, which are stochastic in nature, could take place at different accretion rates with small variations in the disk mass-flow rate or in the magnetic field configuration. This model can reproduce the transitions with comparable torque magnitudes, but has a difficulty in explaining the observed torque reversals with small variation in $\Lx$.         

In the model proposed by \citet{Perna2006}, disk-magnetosphere interaction for an oblique rotator leads to a cyclic variation in the accretion-rate, $\Mdotacc$, while the mass-flow rate from the outer disk is constant. In this model, in the propeller phase, a fraction of the inflowing matter that is ejected from the inner disk returns back to the disk at larger radii, which is the reason for the $\Mdotacc$ variation. Neglecting the magnetic torques, \citet{Perna2006} proposed that the changes in the material torques arising from the accretion/propeller transition due to the changes in $\Mdotacc$ could be responsible for the observed torque reversal. As discussed by these authors, during the torque-reversal, a significant $\Lx$ variation is estimated in the model, which is not in agreement with the observed torque reversal of \src. We do not encounter this luminosity problem in our model, because there is mass accretion onto the star on either side of the torque reversal.

\section{Conclusions}
\label{conc}

We have shown that the torque reversal behavior of \src, the only known likely thin disk, Roche lobe overflow source observed through torque reversals at present, can be reproduced in the comprehensive model developed by \citet{Ertan2021} extending the earlier work on the SP/WP transition \citep{Ertan2017, Ertan2018} to include the torque reversal and the SU phase as well. The model calculates the inner disk radius, torque, and X-ray luminosity as functions of the mass-flow rate for neutron stars accreting from steady, geometrically thin accretion disks. In the model, the observed torque reversals are produced naturally when the system makes transitions between the WP and SU phases with an accretion disk that remains geometrically thin (optically thick) across the torque reversal. We have obtained reasonable model curves with $B \simeq 3 \times 10^{12}$~G and $d \simeq 5.3$~kpc, which produce the torque reversal at an accretion rate $\Mdotstar \simeq 1.9 \times 10^{16}$~\gpers~(Fig. \ref{fig2}). This field strength is in agreement with the $B$ values estimated from the observed cyclotron lines in the $(2.4~-~6.3) \times 10^{12}$~G range \citep{Chakrabarty1998, Orlandini1998, Coburn2002, Dai2017} and the distance is close to the lower edge of the estimated distance range of $5~-~13$~kpc \citep{Chakrabarty1998, Takagi2016}. 

We find that $\rin = \rco$, and $\rxi$ is close to, but not necessarily equal to $\rco$ during the torque reversal. A strong indication of our model results is that both spin-down and spin-up torques are operating on either side of the torque reversal. The spin-down torque produced by the disk-field interaction remains roughly constant during the WP phase and across the torque reversal. With increasing $\Mdotstar$ in the WP phase, the accretion torque gradually dominates the spin-down torque, decreasing the magnitude of the net spin-down torque, and eventually the neutron star makes a transition to the SU phase. The magnitudes of the net spin-down and spin-up torques on either side of the transition are found to be naturally commensurate in this model.

\begin{acknowledgements}
    We thank the anonymous referee, for very useful comments that have considerably improved our manuscript. We also thank K. Yavuz Ek\c{s}i for useful comments on the manuscript. We acknowledge research support from Sabanc{\i} University, and from T\"{U}B\.{I}TAK (The Scientific and Technological Research Council of Turkey) through grant 120F329. A.V. acknowledges support from the Bundesministerium f{\"u}r Wirtschaft und Energie through Deutsches Zentrum f{\"u}r Luft-und Raumfahrt (DLR) under the grant number 50 OR 1917.
\end{acknowledgements}

%
   \bibliographystyle{aa} 
   \bibliography{example} 
%

\end{document}